\documentclass[11pt]{article}
\catcode`@=11
\@addtoreset{equation}{section}
\begin{document}
\begin{center}
{ \Large \bf  Electrostatic acoustic modes in a self-gravitating complex plasma with variable charge impurities}
\end{center}

\vskip 20pt
\begin{center}
{{{\bf {   Amar P. Misra\footnote{apmisra28@rediffmail.com} and A. Roy Chowdhury\footnote{ arcphy@cal2.vsnl.net.in}}\\
                      {\it     High Energy Physics Division,
                             Department of Physics  
                             Jadavpur University,
                              Kolkata - 700 032,
                                    India. \\}}
and\\}
{\bf K. Roy Chowdhury \footnote{kasturi28@yahoo.com}} \\
{\it Department of Physics, J.C.C. College,  Kolkata-700 033, India. }}
\end{center}
\vskip 20pt
\begin{center}
\bf{Abstract}\\
\end{center}
A linear theory of dust acoustic (DA) and dust ion-acoustic (DIA) waves in a self-gravitating collisional dusty plasma contaminated by variable-charge impurities is presented for a self-consistent closed system. The  ion-drag  force arising from the ion orbital motion as well as momentum transfer from all the ions and self-gravitation are taken into consideration.  The physical processes, viz., dust-charge relaxation, ionization, recombination and collisional dissipations are also taken into account self-consistently. The generalized dispersion relation describing the coupling of the acoustic modes with the dust-charge relaxation mode (CRM) and damping of the waves is derived and analyzed numerically.\newpage
{\Large\bf 1. Introduction}\\ \par
Charged impurities or dust grains frequently appear in space and laboratory plasmas [1,2]. Many low-temterature technological plasmas are contaminated by highly charged dust or impurities, since they carry a considerable ammount of negative charge of the plasma. The massive (compared to the ion mass) dust particles significantly affect the charge balance in plasmas, and producing the specific dust-charge fluctuations [3] lead to a new plasma mode, called  charge relaxation mode (CRM). There are many important factors which can significantly affect the dispersion characteristics of dusty plasma waves, for which we need to emphasize the open character of dusty plasma system demanding regorous treatment of collision mechanisms specific to dust-contaminated plasmas, esspecially those leading to plasma recombination.\par In the earlier investigations, it was shown that acoustic modes can be unstable in presence of ion-drag and ionization, and coupling between the DA and DIA modes is strong enough in the long-wave length run (see, e.g., Ref. 4). In the latter, DA and DIA modes interact through the dust-charge branch [5]. This is very different from the direct coupling between the long-wave length acoustic modes without the charge variation. However, in these investigations [4,5] the collisional effects are taken not in a self-consistent manner. Moreover, the ion-drag force which is of practical interest in processing plasmas are also treated as independent on dust charge. In absence of ionization, recombinatiion and collisional dissipations, the Langmuir waves were found to be unstable due to coupling with the CRM [6]. It was also found that the DAWs can be strongly affected by the particle creation and loss mechanisms [7]. The effect of dust-charge variation on Langmuir and ion-acoustic waves in a low-temperature plasma was investigated [8,9] by taking into consideration the self-consistent particle balance. It was shown that dust-charge relaxation is significantly affected by ionization and recombination. These processes maintain the averaged background particle number densities self-consistently during the perturbations by acting as sources and sinks. They also define the equilibrium or steady state. It was also shown that these dissipative processes due to inter-particle collisions as well as elastic Coulomb and the inelastic dust-charging collisions can lead to a net damping of the Langmuir waves [8] as well as waves in typical laboratory plasma [8,9].\par It is known that the ion-drag forces are very important in the dispersion relation for modes in dusty plasma. These are discussed by a number of authors [10,11]. The ion-drag forces may arise from the ion orbital motion around negatively charged dust particles as well as from the momentum transfer from all the ions which are collected by the dust grains. The instability may arise due to the ion fluid compression by the ion-drag force. The latter, which is proportional to the square of the particle radius, dominates over the electrostatic force when the dust grains have sufficiently large sizes.\par The gravitational effects become important when the sizes of the dust grains become considerable. For too small grains, trapping by magnetic field lines dominates much as it does for other plasma particles. Too large grains on the other hand follow gravitationally bound orbits, without being distracted from them by electromagnetic forces. Only for medium-sized grains both the effects balance each other. In plasmas like protostellar clouds, there may be competition between the gravitational self-attraction and electrostatic repulsion between the grains. When self-gravitational interaction due to the heavier dust component is included, dusty plasmas are subject to macroscopic instabilities of the Jeans type. When a relative streaming between the plasma and dust component is present, Jeans-Buneman type instabilities are excited [12-14]. In the absence of equilibrium dust streaming  and the dc electric field, the dust grains are held under the combined influence of the gravity force and the equilibrium drag force involving unperturbed ion flow towards the dust grain surface. \par In this report we reconsider the problem of linear DA and DIA wave propagation in a self-gravitating dusty plasma with dust-charge variations, ionization, recombinations as well as collisional effects due to collisions between the electrons and dust, ion and dust etc. in a self consistent manner.  We also consider the influence of the ion orbital drag force which depends on the ion velocity as well as dust charge. It is found that by these effects the DA and DIA wave propagations as well as instabilities are somewhat modified.\\\\\\
{\Large\bf 2. The Model}\\ \par We consider  the propagation of linear DA and DIA waves in a self-gravitating collisional dusty plasma with finite effective electron $(T_e)$ and  ion $(T_i)$ teperatures. The size of the dust grains$ (r_d)$  is assumed to be much smaller than the average intergrain spacing $(d=(3/4\pi n_{d0})^{1/3}))$, the electron Debye length $(\lambda _{De})$ and the wave length $(r_d<<d<<\lambda_{De},k^{-1})$, so that they can be considered as massive point particles similar to multiply charged negative (or positive) ions in multispecies plasmas [15]. The charge of a dust grain may vary due to microscopic electrons and ion currents flowing into its surface, and these currents appear because of the potential difference  between its surface and the adjacent plasma.  We assume that the ion-drag force $(\vec F_{di})$ arising from the ion orbital motion around the dust grains dominates over that arising from the momentum transfer of all the ions which are collected by the dust grains. This is valid as long as the collection impact parameter is much smaller than the orbital impact parameter and the Coulomb logarithm far exceeds the unity. Thus the orbital motion related ion-drag force acting on the dust grains can be taken as [16] 
$$\vec F_{di} =4\pi n_i m_i V_{it}\vec v_i b^2_{\pi/2} \Lambda(V_{it})  \eqno(1) $$  where $V_{it}=(v^2_i +8V^2_{Ti}/\pi )^{1/2}$ is the total ion speed, $b_{\pi/2}=Z_{d0}e^2/m_i V^2_{it} $ is the orbital impact parameter and $\Lambda(V_{it}) =ln \left[(\lambda ^2_{De}+b^2_{\pi/2})/(b^2_c +b^2_{\pi})\right]^{1/2} $ is the Coulomb logarithm integrated over the interval from the collection impact parameter $b_c=r_d(1+2Z_{do}e^2/r_d m_i V^2_{it}) $ to the electron Debye length. Other parameters will be discussed in the following. \par The dynamics of electrons, ions and negatively charged dust grains in presence of gravity and ion-drag force as well as ionization, recombination, attachments of electrons and ions to the dust, elastic and inelastic collisions and plasma particle collisions is governed by a set of equations:
$$ \frac{\partial n_e}{\partial t}+\nabla.(n_e \vec v_e)=-\nu_{ed}n_e +\nu_{ion}n_e -\beta_{eff}n^2_e  \eqno(2)  $$
$$\frac{\partial \vec v_e}{\partial t}+\nu^{eff}_e \vec v_e +\frac{T_e}{m_e n_e} \frac{\partial n_e }{\partial z} =\frac{e}{m_e} \frac{\partial \phi}{\partial z} \eqno(3) $$
$$ \frac{\partial n_i}{\partial t}+\nabla.(n_i \vec v_i)=-\nu_{id}n_i +\nu_{ion}n_e -\beta_{eff}n^2_e  \eqno(4)  $$
$$\frac{\partial \vec v_i}{\partial t}+\nu^{eff}_i \vec v_i +\frac{T_i}{m_i n_i} \frac{\partial n_i }{\partial z} =\frac{Z_i e}{m_i} \frac{\partial \phi}{\partial z} \eqno(5) $$
 $$ \frac{\partial n_d}{\partial t}+\nabla.(n_d \vec v_d)=0 \eqno(6) $$ 
$$\frac{\partial \vec v_d}{\partial t}+(\vec v{_d}.\nabla)\vec v_d +\nu_{dn} \vec v_d +\frac{T_d}{m_d n_d} \frac{\partial n_d }{\partial z} =-\frac{q_d}{m_d} \frac{\partial \phi}{\partial z}+\frac{\vec F_{di}}{m_d}-\frac{\partial \psi }{\partial z}  \eqno(7) $$
$$\frac{\partial^2  \psi}{\partial z^2}=4\pi Gm_d n_d  \eqno(8)$$
$$\nabla^2 \phi =-4\pi e(Z_i n_i -n_e +\frac{q_d}{e} n_d) \eqno(9) $$
$$\frac{dq_d}{dt}+\nu^{ch}_d q_d =-|I_{e0}| \frac{n_e}{n_{e0}}+|I_{i0}| \frac{n_i}{n_{i0}}  \eqno(10) $$
where $\phi (\psi)$ is the electrostatic (gravitational) potential and $m_j, \vec v_j, n_j$ (including the equilibrium value $n_{j0}$ ) are the mass, velocity and density of the j-th species ( $j=i, e, d$ for ion, electron and dust ) respectively. Furthermore,  $G$ is the gravitational constant, $r_d$ is the dust grain radius and $q_d =-Z_d e$ and $Z_i e$ are the charges of dust components and ions. In Eqs. (2) and (4) $\nu_{(e,i)d}$ are the attachment frequencies of electrons and ions to the dust grain so that $-\nu_{(e,i)d} n_{e,i} $ represent their losses in densities; $\nu_{ion}$ is the rate of electron impact ionization of the neutral atoms, $\beta_{eff}=\beta-\beta_{si}, \beta $ is the volume recombination rate and $\beta_{si}$ is the stepwise ionization rate. In Eqs. (3) and (5) $ \nu^{eff}_{e,i}=\nu_{ei}+ \nu_{(e,i)n}+ \nu^{el}_{e,i}+ \nu^{ch}_{e,i} $ are effective collision frequencies. Here $\nu_{ei}$ is the rate of electron-ion collision, $\nu_{\alpha n}$ is the $\alpha-$particle $(\alpha=e,i)$ collision with neutral, $\nu^{el}_{e,i}$ are the rate of elastic (Coulomb) collisions with the dusts and $\nu^{ch}_{e,i}$ are the inelastic charging collision rates. Also in Eq. (7) $\nu_{dn}$ is the effective frequency of the dust-neutral collision [17].  In the charging  Eq. (10), $\nu^{ch}_{d} =r_d \omega^2_{pi} A/\sqrt{2\pi}V_{Ti} $ is the charging rate of the dust particle [3], defined by the equilibrium electron and ion currents given by 
$$I_{e0}=-\pi r^2_d e \left(\frac{8T_e}{\pi m_e}\right)^{1/2}n_{e0} exp\left(\frac{e\Delta \phi_g}{T_e}\right) \eqno(11)  $$
$$I_{i0}=\pi r^2_d  Z_i e \left(\frac{8T_i}{\pi m_i}\right)^{1/2}n_{i0} \left(1-\frac{e\Delta \phi_g}{T_i}\right) \eqno(12)  $$
where $\Delta \phi _g =q_{d0}/C, C=r_d (1+r_d /\lambda _{De}) $ is effective grain capacitance and $q_{d0}$ the stationary grain charge. Here $A=1+\tau +Z, \tau =T_i /T_e, Z=z_d e^2 /r_d T_e, V_{T(e,i)}=\sqrt{(T_{e,i}/m_{e,i}}),     \omega_{p(i,d)}=\sqrt({4\pi n_{(i,d)0}Z^2_{i,d} e^2 /m_{i,d}}). $ The rates of elastic electron and ion-dust Coulomb collisions are 
$$\nu^{el}_{e}=\frac2 3 \nu^{ch}_d \frac{P(A-1)\Lambda}{AZ}exp(Z)=\nu^{el}_i \frac{n_{i0}}{n_{e0}}\tau (A-1)exp(-Z) \eqno(13) $$ where $\Lambda$ is the Coulomb logarithm. The effective charging rates are [18] 
$$\nu^{ch}_e =\frac{P(\tau +Z)(4+Z)}{AZ}=\frac3 2 \nu^{ch}_i \frac{n_{i0}}{n_{e0}}\frac{(\tau +Z)(4+Z)}{(2\tau +Z)} \eqno(14) $$ where $P=Z_d n_{d0}/n_{e0}$.  Also the rates of electron and ion capture at the grain surface is [3,18]
$$\nu_{ed}=\frac{P(\tau +Z)}{ZA}\nu^{ch}_d =(1+P)\nu_{id} \eqno(15)  $$ The expressions for the ionization, recombination rates as well as for electron-ion, ion-neutral collisions etc. can be had from Ref. [19].\\\\\\
{\Large\bf 3. Dispersion relation}\\\par To obtain the self-consistent equilibrium state we assume that the pressure is not  so low such that the recombination losses prevail over the diffusion losses, i.e., ambipolar diffusion can be neglected. In equilibrium, the system is quasi-neutral so that 
$$Z_{i0}n_{i0}=n_{e0}+Z_{d0}n_{d0} \eqno(16) $$ From Eqs. (2) and (4) we obtain the stationary electron and ion densities as
$$n_{e0}=\frac{\nu_{ion}-\nu_{ed}}{\beta_{eff}}, \hskip 50 pt  n_{i0}=\frac{\nu_{ed}}{\nu_{id}}n_{e0}  \eqno(17)  $$
 We note that the ionization degree must be high enough so that $\nu_{ion}>\nu_{ed}$, otherwise no self-consistent stationary state exists.\par Assuming the perturbations to vary as $exp(ikz-i\omega t) $ and linearizing  the Eqs. from (2)--(10) we obtain for the perturbed electron density and velocity $(n_j=n_{j0}+n_{j1},   j=e,i;  Z_d=Z_{d0}+Z_{d1} )$ 
$$n_{e1}=\frac{in_{e0}}{\eta_e}\left[\frac{iek^2 \phi}{m_e (\omega+i\nu^{eff}_e)}-\left(\frac{\partial \nu_{ed}}{\partial z_d}\right)_{Z_{d0}} Z_{d1}\right]  \eqno(18) $$
$$v_e=\frac{k\left(\frac{V^2_{Te}}{n_{e0}}n_{e1}-\frac e{m_e}\phi\right)}{\omega+i\nu^{eff}_e} \eqno(19)  $$where $\eta_e=\omega-i(\nu_{ed}-\xi _2\nu_{ion})-k^2 V^2_{Te}/(\omega+i\nu^{eff}_e),  \xi_2=2-\xi_1.$  Here $\xi_1$ depends on the used model of the direct ionization. $\xi_1 =0 $ corresponds to density-fluctuation independent ionization, whereas $\xi_1=1$ that for density-fluctuation dependent.\par For the ion fluid velocity and perturbed ion density we have $$n_{i1}=-\frac{n_{io}}{\eta_i}(iek^2 \phi \Lambda_1 +Z_{d1}\Lambda_2 ) \eqno(20) $$
$$v_i=\frac{k\left(\frac{V^2_{Ti}}{n_{i0}}n_{i1}-\frac{ Z_i e}{m_i}\phi\right)}{\omega+i\nu^{eff}_i} \eqno(21)  $$where  
$$\eta_i=\omega-i\nu_{id}-\frac{k^2 V^2_{Ti}}{\omega+i\nu^{eff}_i },\hskip 20 pt  \zeta =\frac{n_{e0}}{n_{i0}}\frac{\xi_2 \nu_{ion}-2\nu_{ed}}{\eta_e}  $$
$$\Lambda_1=\frac{Z_i}{m_i(\omega+i\nu^{eff}_i)}+\frac{i\zeta}{\eta_e m_e(\omega+i\nu^{eff}_e)}, \hskip 10pt \Lambda_2=\left(\frac{\partial \nu_{id}}{\partial Z_d}\right)_{Z_{d0}}-i\frac{\zeta}{\eta_e}\left(\frac{\partial \nu_{ed}}{\partial Z_d}\right)_{Z_{d0}} $$ For the dust charge and number density perturbation we have 
$$Z_{d1}=-i\frac{|I_{e0}| k^2 \phi \Lambda_3}{\eta_e(\omega+i\widetilde{\nu}^{ch}_d)} \eqno(22)  $$
$$n_{d1}=\frac{n_{d0}k}{\Gamma m_d}\left[q_{d0}k\phi +i8(2\pi)^{1/2}n_{i0}m_i V_{Ti}b^2_{id}\Lambda_0 v_{i}\right] \eqno(23) $$ Here $$\Lambda_3=\frac{1+i\zeta/\eta_i}{m_e(\omega+i\nu^{eff}_e)}+\frac{i\eta_e Z_i}{\eta_i m_i(\omega+i\nu^{eff}_i)} $$
$$\widetilde{\nu}^{ch}_d=\nu^{ch}_d +\frac{i|I_{e0}| }{e\eta_e}\left[(1+i\zeta/\eta_i)\left(\frac{\partial \nu_{ed}}{\partial Z_d}\right)_{Z_{d0}}-\frac{\eta_e}{\eta_i}\left(\frac{\partial \nu_{id}}{\partial Z_d}\right)_{Z_{d0}}\right] \eqno(24)  $$ Also 
$\Gamma=\omega^2 +i\omega \nu_{dn}-k^2 V^2_{Td}-\omega^2_{Jd},  V^2_{Td}=T_d/m_d ,  b_{id}\approx \pi Z_{d0} e^2/8T_i ,  \Lambda_0=\Lambda(V_{it0})$ and $\omega^2_{jd} =4\pi Gm_d n_{d0}$ is the Jeans' frequency.  Finally, substitutions of Eqs. (18)--(23) into Eq. (9) lead to the dispersion relation for DA and DIA waves
$$\epsilon _{pd}=\frac{\omega^2_{pe}}{\omega+i\nu^{eff}_e}\zeta_e +\frac{\omega^2_{pi}}{\omega+i\nu^{eff}_i}\zeta_i \eqno(25) $$ where $$\zeta_e=\frac{k^2}{\eta_e}\left[1+i\frac{Z_i}{\eta_i}(\xi_2 \nu_{ion}-2\nu_{ed})\right], \hskip 20 pt  \zeta_i=k^2/\eta_i $$ 
$$\epsilon _{pd}=k^2 +\frac{\widetilde{\eta} m_d \omega^2_{pd}}{Z_{d0}e \Gamma}+i\frac{4\pi e|I_{e0}| k^2 \Lambda_3}{\eta_e(\omega+i\widetilde{\nu}^{ch}_d)}\left[i\frac{n_{e0}\zeta_e}{k^2}\left(\frac{\partial \nu_{ed}}{\partial Z_d}\right)_{Z_{d0}} -i\frac{Z_i n_{i0}}{\eta_i}\left(\frac{\partial \nu_{id}}{\partial Z_d}\right)_{Z_{d0}}-n_{d0}\right] \eqno(26) $$
$$\widetilde{\eta}=\frac{k^{2}}{m_d}\left[q_{d0} +i8(2\pi)^{1/2}n_{i0}m_i \frac{ V_{Ti}b^2_{id}\Lambda_0}{\omega+i\nu^{eff}_i}\left(\frac{eZ_i}{m_i}+\frac{k^2 V^2_{Ti}}{\eta_i}\left(e\Lambda_1 - \frac{|I_{e0}|\Lambda_2 \Lambda_3}{\eta_i(\omega+i\widetilde{\nu}^{ch}_d)}\right)\right)\right] \eqno(27)$$ 
Equation (25) describes the linear coupling of Langmuir, DA and DIA waves with the CRM $\omega=-i\nu^{ch}_d$ where the effects of the variation of electron and ion capture rate by the dusts are taken into consideration. Since it is very difficult to analyze Eq. (25) either analytically  or numerically keeping all the terms in order, we rewrite the equation in a comparatively simplified form on the assumption that
$$\left(\frac{\partial \nu_{e,i}}{\partial Z_d}\right)_{Z_{d0}}<<\nu_{(e,i)d} \eqno(28) $$ However, for larger dust grain size and higher ion number density, the condition (28) can be violated. Note that one can recover the dispersion relation equation of Ref. [8] for Langmuir plasma waves. Here we only discuss the DA and DIA modes. Thus, neglecting the the variations of $\nu_{(e,i)d}$ with respect to $Z_d$ we obtain from (25) 
$$D(\omega,k) D_{ch}(\omega)=i\beta_{coupl}(\omega,k) \eqno(29)  $$
Here $$D(\omega,k)=1-\frac{\omega^2_{pe}}{\Lambda_e}-\frac{\omega^2_{pi}}{\Lambda_i}-\frac{\omega^2_{pd}}{\Lambda_d}, \hskip 10 pt  \Lambda_e=\chi  /(\omega^2 +i\Gamma_3 \omega +\zeta_3 -k^2 V^2_{Ti}) $$
$$\chi =(\omega^2 +i\Gamma_1 \omega +\zeta_1 -k^2 V^2_{Te})\Lambda_i, \hskip 10 pt \Lambda_i=
(\omega^2 +i\Gamma_2 \omega +\zeta_2 -k^2 V^2_{Ti}) $$
$$\Lambda_d=\frac1 \Gamma \left[1-i\frac{\omega^2_{pd}\omega_{id}}{\omega+i\nu^{eff}_i}\left(1+\frac{k^2 V^2_{Ti}}{\chi }(\omega^2 +i\Gamma_4 \omega +\zeta_4 -k^2 V^2_{Te})\right)\right]  $$
$$\zeta_1=\nu^{eff}_e(2\nu_{ed}-\xi _2\nu_{ion}), \hskip 10 pt \zeta_2=\nu_{id}\nu^{eff}_i $$
$$\zeta_3=\zeta_2+Z_i \zeta_1\nu^{eff}_i/\nu^{eff}_i, \hskip 5pt \zeta_4=\zeta_2+\frac{\nu_{id}}{\nu_{ed}}\zeta_1\frac{\nu^{eff}_i}{\nu^{eff}_i}$$
$$\Gamma_1=\nu^{eff}_e -2\nu_{ed}+\xi _2 \nu_{ion}, \hskip 10pt \Gamma_2= \nu^{eff}_i -\nu_{id}, \hskip 10pt \Gamma_3=\Gamma_2+Z_i(\xi _2 \nu_{ion}-2\nu_{ed}) $$
$$\Gamma_4=\Gamma_1+\frac{m_i}{m_e}\frac{\nu_{id}}{Z_i \nu_{ed}}(\xi _2 \nu_{ion}-2\nu_{ed}), \hskip 20pt D_{ch}(\omega)= \omega+i\nu^{ch}_d  $$
$$\beta_{coupl}(\omega,k)=\frac{\widetilde{\nu}}{\chi }(\omega^2 +i\Gamma_5 \omega +\zeta_5 -k^2 V^2_{Ti}-\frac{m_e}{m_i}Z_i k^2 V^2_{Te})\omega^2_{pe} $$
$$\Gamma_5=\Gamma_4+\frac{m_e}{m_i}Z_i \Gamma_1, \hskip 10 pt \zeta_5=\zeta_4 -\frac{m_e}{m_i}Z_i \zeta_1, \hskip 10pt \widetilde{\nu}=\frac{|I_{e0}|n_{d0}}{en_{e0}} $$ Note that $\widetilde{\nu}$ is the correction of the dust charge relaxation frequency arising from the perturbation of the electron density. \par Equation (29) describes the coupling of the acoustic waves given by $D(\omega,k)=0$ with the CRM given by $D_{ch}(\omega)=0 $. We discuss the two modes separately as below by considering the hierarchy of the frequency scales for the DIA and DA respectively as $|\omega_{DIA}|\sim 10^5-10^7/s$ and $|\omega_{DA}|\sim 1-10^2/s.$\\\\
{\large\bf I. Dust ion-acoustic modes}\\ \par Assuming $ kV_{Ti},kV_{Td}<<\omega<<kV_{Te}$ and $T_d<<T_e, T_i $ we obtain from Eq. (29) the following dispersion relation for DIA modes$$\left[\omega^2 D_{iff}\left(\omega^2(k^2 V^2_{te}+\omega^2_{pe})+i\omega(\Gamma_2 k^2 V^2_{Te}+\Gamma_3 \omega^2_{pe})+(\zeta_2-\omega^2_{pi})k^2 V^2_{Te}+\zeta_3 \omega^2_{pe}\right)\right]D_{ch}+$$
$$\left[ik^2 V^2_{Te}\omega^2_{pd}\omega_{id}(\omega^2+i\Gamma_2 \omega+\zeta_2)\right]D_{ch}+i\omega^2\omega^2_{pe}\widetilde{\nu}D_{iff}(\omega^2 +i\Gamma_5 \omega +\zeta_5 -\frac{m_e}{m_i}Z_i k^2 V^2_{Te}) =0 \eqno(30) $$where $D_{iff}=(\omega+i\nu^{eff}_i)$. 
Setting $\omega= \omega_1 +i\delta^{\prime}_1+i \delta^{\prime \prime}_1 $, where $\omega_1=\omega_{pi} k\lambda_{De}$, we find from (30) in the limit $k^2 \lambda^2_{De} <<1 $ 
$$\delta^{\prime}_1=-\frac{QR_1 +PR_2}{P^2 +Q^2} \eqno(31) $$
$$\delta^{\prime \prime}_1=-\frac{QR_2 -PR_1}{P^2 +Q^2} \eqno(32) $$
where $$ P= 5\omega ^4_1 a_1-3\omega ^2_1 a_3+a_5, \hskip 10pt Q= 6\omega ^5_1 -4\omega ^3_1 a_3+2\omega_1 a_4 $$
$$R_1=\omega ^6_1 -\omega ^4_1 a_2 +\omega ^2_1 a_4 -a_6 , \hskip 10pt  R_2= a_1\omega ^5_1 -\omega ^3_1 a_3+\omega_1 a_5$$
$$a_1=-(\widetilde{\nu}+\nu^{ch}_d+\nu^{eff}_i+\Gamma_3), \hskip 5pt a_2=-\zeta_3 +\widetilde{\nu}(\Gamma_5+\nu^{eff}_i)+\nu^{ch}_d(\nu^{eff}_i+\Gamma_3)+\nu^{eff}_i \Gamma_3 +k^2 \lambda^2_{De}\omega^2_{pi} $$
$$a_3=\widetilde{\nu}(\zeta_5-\nu^{eff}_i \Gamma_5)+\nu^{ch}_d(\zeta_3 -\Gamma_3 \nu^{eff}_i)+\zeta_3 \nu^{eff}_i -k^2 \lambda^2_{De}\left(\omega^2_{pi}(\nu^{eff}_i +\nu^{ch}_d)-\omega^2_{pd} \omega_{id}\right) $$
$$a_4=-(\widetilde{\nu}\zeta_5 +\nu^{ch}_d \zeta_3)\nu^{eff}_i -k^2 \lambda^2_{De}\left[\nu^{ch}_d(\omega^2_{pd}\omega_{id}-\omega^2_{pi} \nu^{eff}_i)+\omega^2_{pd}\omega_{id}\Gamma_2\right] $$
$$a_5 =k^2 \lambda^2_{De}\omega^2_{pd}\omega_{id}(\nu^{ch}_d \Gamma_2-\zeta_2), \hskip 20 pt a_6 =k^2 \lambda^2_{De}\omega^2_{pd}\omega_{id}\nu^{ch}_d \zeta_2$$ In Eq. (31) negative sign indicates that a frequency downshift takes place for the case considered, and comparing the values of all dissipative terms one can show from Eq. (32) that in presence of ion-drag force there may be an increase in the damping decrement. In a dusty argon plasma with $T_e\sim 10$ eV, $T_i \sim 1$ eV, $T_n\sim 0.015$ eV, $n_{io}/n_{e0}\sim 10, r_d\sim 5\mu $m, $Z_{d0}\sim 10^4 $ we  find for $\xi_2=2$, $\delta^{\prime}_1\approx -2.393\times10^6, \delta^{\prime \prime}_1\approx -1.54 \times10^6 $ in absence of ion-drag and $\delta^{\prime}_1\approx -2.3545\times10^6/s, \delta^{\prime
 \prime}_1\approx -1.5737 \times10^6/s $ in presence of ion-drag. That is the ion-drag force leads to decreasing of the amount of shift and the growth rate of damping. From the detailed numerical analysis  we also find that the amount of shift increases with the electron-to-ion temperature as well as ion-to-electron density ratio. \par The Eq. (28) is higher order in the wave eigen frequency, for experimental practice one can at once find the spatial attenuations by setting $k\lambda_{De}=K_1+iK_2$ as 
$$ K^2_{1,2}=\frac{\pm(X_D X_N+Y_D Y_N)+\sqrt{(X^2_D +Y^2_D)(X^2_N +Y^2_N)}}{2(X^2_D +Y^2_D)} \eqno(33)$$
Here $$X_D=\omega^2\chi _1-\nu^{ch}_d \chi _2+\Delta \nu^{eff}_{i}, \hskip 20 pt  X_N=\omega^2(\omega^2 \chi_3 -\nu^{eff}_i\chi _4)$$
$$Y_D=\omega(\nu^{ch}_d \chi _1+\chi _2 -\Delta), \hskip 5pt Y_N=\omega^3(\nu^{eff}_i \chi _3 +\chi _4), \hskip 5pt \Delta =\frac{m_e}{m_i}Z_i \widetilde{\nu}\omega^2_{pe} \omega^2 $$
$$\chi _1=\omega^2(\omega^2+\zeta_2-\omega^2_{pi})-\omega^2\Gamma_2\nu^{eff}_i-\omega^2_{pd}\omega_{id}\Gamma_2$$
$$\chi _2=\omega^4 \Gamma_2+\omega^2\nu^{eff}_i(\omega^2+\zeta_2-\omega^2_{pi})+\omega^2_{pd}\omega_{id}(\omega^2+\zeta_2)$$
$$\chi _3=\omega^2(\omega^2+\zeta_3-\nu^{ch}_d\Gamma_3-\widetilde{\nu}\Gamma_5), \hskip 20pt
\chi _4=\nu^{ch}_d(\omega^2+\zeta_3)+\widetilde{\nu}(\omega^2+\zeta_5)$$For a typical dusty plasma the DIA wave frequency is $\sim 10^6/s$, so that the self-gravitational effect on DIA modes being smaller can be neglected.
We have plotted $K_1$ and $K_2$ vs real $\omega$ for different values of electron to ion temperature and ion to electron density ratio (Figs.(1), (2)). We find that both the ratios increase the damping decrement of the wave. But when  the density ratio takes the value 6.0  the propagation stops in a short range of frequency domain and the maximum decay rate is then $\approx 4.6\times10^{-2}$ (Fig. 2 dashed line ). Further increase of ion density increases the damping decrement.  We numerically verify the condition $|k^2\lambda^2_{De}|<<1$ and the restriction $\omega<\omega_{pi}$ for DIA modes. To estimate the ion number density we find that for a typical argon plasma parameters described above the threshold ion number density $(n^{thres}_{i0})$ is $\sim 9.0\times10^{11}cm^{-3}$               when $\omega\approx 1.07\times10^8/s, K_1\approx 0.291, K_2\approx 0.953.$ A slight decrease of $n_{i0}=8.95\times10^{11}cm^{-3}$ would lead to a increase of $K_1$ and $K_2$ $(K_1\approx 0.293, K_2\approx 0.958 )$ and hence a violation of the condition for $k\lambda_{De}.$\\\\
{\large\bf II.Dust acoustic modes}\\\par Assuming $kV_{Td}<<\omega<<kV_{Te},KV_{Ti}$ we obtain from Eq. (29) the following dispersion relation for DA modes
$$\left[D_{iff}D_{jd}(i\Gamma_3\omega+\zeta_3-k^2 V^2_{Ti}-k^2 \lambda^2_{De}\omega^2_{pi})+i\omega^2_{pd}\omega_{id}k^2\lambda^2_{De}\left(i\omega(\nu^{eff}_i-\eta\omega_{id})+\nu_{id}\nu^{eff}_i\eta \right)\right]D_{ch}$$
$$+\widetilde{\nu}(\omega+i\nu^{eff}_i)(i\Gamma_5\omega+\zeta_5-k^2 V^2_{Ti}-\frac{m_e}{m_i}Z_i k^2 V^2_{Te})D_{jd}=0 \eqno(34)$$where $$D_{jd}=(\omega^2+i\omega\nu_{dn}-\omega^2_{jd}), \hskip 5pt \eta=1+\tau \frac{\xi_2\nu_{ion}-2\nu_{ed}}{Z_i \nu_{ed}} $$ A detailed numerical analysis shows that frequency downshift occurs due to coupling of DA modes with the CRM. \par From Eq. (34) we obtain for $k\lambda_{De}=K_R+iK_I$ as 
$$K_R\approx \sqrt{\frac{\widetilde{X}_N}{\widetilde{X}_D}} \eqno(35) $$
$$K_I \approx \frac{\omega}{\widetilde{X}_D}\sqrt{\widetilde{Y}_D \widetilde{Y}_N} \eqno(36) $$
Here$$ \widetilde{X}_D=\omega^2_{pi}\left[\omega^2 \widetilde{\eta}\Delta_1-\Delta_2\Delta _3\right]
+\omega^2_{pd}\omega_{id}\left(\omega^2(\nu^{eff}_i-\eta\omega_{id})+\nu^{ch}_d\nu_{id}\nu^{eff}_i \eta \right) $$
$$\widetilde{Y}_D=\omega^2_{pd}\omega_{id}\left(\nu_{id}\nu^{eff}_i\eta -\nu^{ch}_d(\nu^{eff}_i-\eta\nu_{id})\right)+\omega^2_{pi}(\widetilde{\eta}\Delta _2 +\Delta _1\Delta _3) $$
$$\widetilde{X}_N=\omega^2 \Delta _1\Delta _4 -\Delta_2 \Delta_5 , \hskip 10pt
\widetilde{Y}_N=\Delta _1\Delta _5+ \Delta _4 \Delta _2 $$
$$\widetilde{\eta}=1+\tau \frac{\nu_{id}}{Z^2_i \nu_{ed}},\hskip 10pt
\Delta _1=\omega^2-\omega^2_{jd}-\nu^{eff}_i \nu_{dn}$$
$$\Delta _2=\omega^2 \nu_{dn}+\nu^{eff}_i(\omega^2-\omega^2_{jd}),\hskip 10pt
\Delta _3=\nu^{ch}_d \widetilde{\eta}+\widetilde{\nu}(1+\nu_{id}/ Z_i \nu_{ed})$$
$$\Delta _4=\zeta_3-\nu^{ch}_d\Gamma_3-\widetilde{\nu}\Gamma_5 , \hskip 10pt
\Delta _5=\zeta_3 \nu^{ch}_d +\widetilde{\nu}\zeta_5+\omega^2 \Gamma_3$$
Note that in the dispersion relation (34) we have included the self-gravitation due to the heavier dust components, and in the derivation of Eqs. (35) and (36) we assumed that $\widetilde{X}_{D,N}>>\widetilde{Y}_{D,N} $. For an argon plasma described earlier  we find that $\nu^{eff}_{e,i}\sim 10^5/s, \nu_{dn}\sim 10^{-6}/s, \nu_{ion}\sim 10^6/s $ so that the above approximation can be made. \par From Eq. (34) we find that $\widetilde{X}_{D}>$ or $<0$ according as 
$$\omega^2 < or > \frac{\omega_{id}\omega^2_{pd}\nu^{ch}_d \nu_{id}+\omega^2_{jd}\omega^2_{pi}\Delta_3}{\omega^2_{pi}\Delta _3 (\nu_{dn}+\nu^{eff}_i )}\nu^{eff}_i \eqno(37) $$
This shows that both the gravity and ion-drag force stabilize the instability of the DA modes.
The numerical calculations for (35) and (36) are carried out for the parameters of a dusty argon plasma and plotted against real $\omega$ as shown in the Figs. (3)-(5). To show clearly the influence of gravity on the DA modes we numerically calculate an estimate for $n_{d0}$ in presence and absence of self-gravity to verify the restrictions $\omega<\omega_{pd}$ and $|k^2\lambda^2_{De}|<<1.$ We find that for the parameters $n_{i0}/n_{e0}=10.0, T_e/T_i=10.0, r_d\sim 5\mu m, n_{n0}\sim 3.0\times10^{14}cm^{-3}$ the threshold for $n_{d0}$ is $\approx 13.92\times10^7cm^{-3}$ when $\omega\approx 0.335/s, K_1\approx 1.83\times10^{-5}, K_2\approx -5.89\times10^{-6}$ below which the DA modes do not exist in presence of gravity. Whereas in absence of gravity the upper limit for $n_{d0}$ for the existence of DA modes is $\approx 13.05\times10^7cm^{-3}.$ We also find that in presence of gravity as the dust number density increases from the threshold value, the instability of the modes tend to decrease (Fig. 4, the solid, dotted, dashed and dadotted lines), e.g., at a particular frequency $\omega=0.24/s,$ the values of $K_2$ are $ -5.48\times10^{-5}, -4.88\times10^{-5}, -3.79\times10^{-5}$ and $-1.22\times10^{-5}$ for $n_{d0}=13.9\times10^7cm^{-3}, 15.7\times10^7cm^{-3}, 17.4\times10^7cm^{-3}$ and $19.1\times10^7cm^{-3}$ respectively. That is, gravity influences to stabilize the instability in  the dense region of dust grains.  Figure (3) shows that for a particular dust density where $\omega^2_{jd}=7.29\times10^{-2}s^{-2},$ if we increase the value of $T_e/T_i$ from 1.2  the value of $K_I(>0)$  also increases, i.e.,  electron and ion temperature also play an important role to the damping decrements of the modes in presence of gravity.  We have also shown that for  particular values of temperatures and densities $(T_e/T_i=1.6, n_{d0}=8.7\times10^7cm^{-3}, n_{i0}/n_{e0}=10.0), K_I$ again becomes negative but in absence of gravity and its magnitude increases with the frequency (Fig. 5). \\\\
{\Large\bf Conclusions}\\ \par We have presented a general model for linear acoustic waves
 as well as Langmuir waves self-consistently with the presence of dust grain charge
 variation as well as the influence of ion-drag and self-gravitation. The model also
 includes the ionization, volume recombination, electron and ion capture by the dusts, as
 well as Coulomb elastic and inelastic charging collisions of the particles. From our
 general dispersion relation one can recover Langmuir wave as in Ref. [8], DA and DIA waves
 as in Ref. [4,5] in some particular considerations and also the DA waves as in Ref. [7] for
 stationary neutrals. We have shown that in presence of ionization, recombination and all
 other collisions, linear coupling of both DA and DIA waves with the dust charge mode (CRM)
 leads to damping and a frequency downshift of the waves. This is typical consequence of the
  coexistence of acoustic modes and CRM in dust-contaminated plasmas [9]. The presence of ion-drag force which is opposite to the electrostatic force decreases the amount of shift and the damping increment of the waves. We have presented graphically the dependence of real and imaginary part of DA and DIA wave number on the wave eigen frequency for a dusty argon plasma for different values of temperatures and densities.  We have  shown
  that the instability of the acoustic modes are modified somewhat by the  influence of
  ion-drag and self-gravity force. From our numerical analysis we have estimated the ion and dust number densities due to the restrictions $\omega<\omega_{pi}$ and $\omega<\omega_{pd}$ for DIA and DA modes, and found that in the region of dense dust grains the gravity influences to stabilize the instability of the DA modes. Due to our lack of experimental data we could not compare our results with those of existing  experiments.   Our results will be more applicable to many laboratory as
  well as astrophysical plasmas.\\\\
{\Large\bf Acknowledgements}\\\par It is a great pleasure for A. P. M. to thank the Council of Scientific and Industrial Research (Govt. of India)  for a research fellowship. K. R. C. is thankful to University Grants Commission (India) for a minor research project. Valuable  comments from the Referee for the improvement of the work  are gratefully acknowledged.
\newpage
{\Large\bf References}\\\\
{[1]} C. K. Goertz,  Rev. Geophys. 27 (1989) 271.\\
{[2]} A. Bouchoule and L. Boufendi,  Plasma Sources Sci. Technol.  3 (1994) 262.\\
{[3]} V. N. Tsytovich and O. Havnes:  Comments Plasma Phys. Cotr. Fusion 15 (1993) 267.\\
{[4]} A. V. Ivlev, D. Samsonov, J. Goree et al  Phys. Plasmas 6 (1999) 741.\\
{[5]} A. V. Ivlev and G. Morfill,  Phys. Plasmas  7 (2000) 1094.\\
{[6]} J. X. Ma and M. Y. Yu,  Phys. Rev. E 50 (1994) R2431.\\
{[7]}  K. N.  Ostrikov, S. V. Vladimirov, M. Y. Yu and L. Stenflo,  Phys. Plasmas 7 (2000) 250.\\
{[8]} S. V. Vladimirov, K. N. Ostrokov, M. Y. Yu and L. Stenflo, Phys.  Rev. E  58 (1998) 6578.\\
{[9]}  S. V. Vladimirov, K. N. Ostrokov and M. Y. Yu, Phys.  Rev. E  60 (1999) 3257.\\
{[10]} T. Nitter,  Plasma Sources Sci. Technol.,  5 (1996) 93.\\
{[11]} J. P. Boeuf and C. Punset, {\it Dusty Plasmas: Physics, Chemistry and Technological impacts in Plasma Processing}  ed A. Bouchoule (New York: Wiley) pp-1-76 (1999).\\ 
{[12]} K. Avinash and P. K. Shukla, Phys. Lett. A 189 (1994) 470.\\
{[13]} P. Meuris, F. Verheest and G. S. Lakhina, Planet Space Sci. 45 (1997) 449.\\
{[14]} F. Verheest, P. K. Shukla, N. N. Rao et al, J. Plasma Phys. 58 (1997) 163.\\
{[15]} V. N. Tsytovich and K. Watanabe  Contribution Plasma Phys. 43 (2002) 51;\\
V. N. Tsytovich and K. Watanabe, {\it Universal Instability of Dust Ion-Sound Waves and Dust-Acoustic Waves }(preprint), NIFS-720, jan. 2002.\\  
{[16]} P. K. Shukla and A. A. Mamun, {\it Introduction to Dusty Plasma Physics} ( Instt. of Phys. Publishing Ltd., Bristol, 2002).\\
{[17]} N. D' Angelo, Phys. Plasmas, 5 (1998) 3155.\\
{[18]} S. V. Vladimirov and V. N. Tsytovich, Phys. Rev. E 58 (1998) 2415.\\
{[19]} E. W. McDaniel, {\it Collision Phenomena in Ionized Gases }(Wiley,  New York, 1964); L. M. Biberman, V. S. Vorob'ev and I. T. Yakubov, {\it Kinetics of Non-Equilibrium Low-Temperature Plasmas} (Nauka, Moscow, 1982).

\newpage {\Large\bf Figure Captions}\\\\
Figure (1): Plot of the real (a) and the imaginary (b) parts of $k \lambda_{De}$ vs the DIA wave frequency $\omega \sim10^6/s $ (Eq.(30)) in an argon plasma. The solid $(0\leq \omega \leq 118.72 )$, dottted $(0\leq \omega \leq 119.44 )$, dashed $(0\leq \omega \leq 116.96 )$ and dadotted $(0\leq \omega \leq 118.6 )$ lines correspond to $T_e/T_i =1.0,  3.13,  5.0$  and  19.2  respectively. The other parameter values are given in the text. The corresponding frequency intervals indicating those where the waves exist. For $\omega=0,  K_1=0$ and $K_2\approx 0.$\\\\
 Figure (2): Plot of the real (a) and the imaginary (b) parts of $k \lambda_{De}$ vs the DIA wave frequency $\omega \sim10^6/s $ (Eq.(30)) in an argon plasma. The solid $(0<\omega \leq 77.0 )$, dottted $(0<\omega \leq 164.8 )$, dashed $(0< \omega \leq 2.8 )$ and dadotted $(0<\omega \leq 236.0 )$ lines correspond to $n_{i0}/n_{e0} =1.0,  4.0,  6.0$ and  8.0  respectively. The other parameter values are given in the text. The corresponding frequency intervals indicating those where the waves exist. As $\omega\rightarrow 0, K_1, K_2\rightarrow 0.$\\\\
Figure (3): Plot of the real (a) and the imaginary (b) parts of $k \lambda_{De}$ vs the DA wave frequency $\omega \sim1/s$ (Eq.(34)) in an argon plasma. The solid $(0\leq \omega \leq 0.21 )$, dottted $(0\leq \omega \leq 0.23 )$ and dashed $(0\leq \omega \leq 0.24 )$  lines correspond to $T_e/T_i =1.2, 1.6$ and 2.0  respectively.  Here $\omega^2_{jd}=7.29\times 10^{-2}/s^2$ and the other parameter values are given in the text. The corresponding frequency intervals indicating those where the waves exist. For $\omega=0,  K_1\approx 10^{-3}$ and $K_2= 0.$\\\\
Figure (4): The influence of gravity on DA modes: \\Plot of the real (a) and the imaginary (b) parts of $k \lambda_{De}$ vs the DA wave frequency $\omega \sim1/s$ (Eq.(34)) in an argon plasma. The solid $(0\leq \omega \leq 0.335 )$, dottted $(0\leq \omega \leq 0.332 )$, dashed $(0\leq \omega \leq 0.314 )$ and dadotted $(0\leq \omega \leq 0.249 )$ lines correspond to $\omega^2_{jd} =0.1166, 0.1312, 0.1458$ and  $0.1604/s^2$ respectively. The two lower solid lines of both (a) and (b) are corresponding to no self-gravitation. The other parameter values are given in the text. The corresponding frequency intervals indicating those where the waves exist. For $\omega=0,  K_1$ and $K_2\approx 0.$\\\\
Figure (5):  The same as the dotted line in Fig. (3). Here $K_I$ becomes negative in absence of gravity, showing that instability increases with the frequency.

\end{document}